\documentclass[a4paper,11pt]{article}
\pdfoutput=1 

\usepackage{jheppub} 

\usepackage[T1]{fontenc} 

\def\ARR{A_{1,\frac{1}{2}}}
\def\AZR{A_{0,\frac{1}{2}}}
\def\AZL{A_{0,-\frac{1}{2}}}
\def\ALL{A_{-1,-\frac{1}{2}}}
\newcommand{\Ys}[2]{Y_{#1}^{#2}(\theta^{*},\phi^{*})}

\newcommand{\Pt}[2]{P_{#1}^{#2}(\cos{\theta})}

\title{\boldmath Single top quark differential decay rate formulae including detector effects}

\author{J. Boudreau$^{1}$,\note{Corresponding author.}}
\author{C. Escobar,}
\author{J. Mueller,}
\author{K. Sapp,}
\author{J. Su}
\affiliation{Pittsburgh Particle Physics Astrophysics and Cosmology Center (PITT PACC)\\
Department of Physics and Astronomy, University of Pittsburgh, Pittsburgh PA 15260, USA}

\emailAdd{boudreau@pitt.edu}
\emailAdd{mueller@pitt.edu}
\emailAdd{Carlos.Escobar@cern.ch}
\emailAdd{Kevin.Sapp@cern.ch}
\emailAdd{Jun.Su@cern.ch}

\abstract{Since the discovery of parity violation in 1957, angular 
distributions of leptons coming from the weak decay of polarized fermions
have been used to probe the structure of the $Wqq^\prime$ vertex.  Vector 
and axial vector couplings reveal themselves in the angular distributions 
of both light and heavy polarized fermions, but tensor and pseudotensor 
couplings have a prominent influence on the angular distributions only
for fermions heavier than the $W$ boson; i.e. the top quark.  The copious 
$t$-channel production of polarized single 
top quarks at the LHC provides an opportunity to study the angular 
distributions of leptons from polarized top quark decay.  In this paper
we develop formulae for differential rates intended to be used as a 
likelihood function in the simultaneous extraction of decay amplitudes,
phases, and polarization.  The incorporation of detector effects in these
formulae is accomplished using a variant of the familiar convolution
theorem applying to a decomposition of the differential rates
in spherical harmonics.}

\begin{document} 
\maketitle
\flushbottom

\section{Introduction}
\begin{figure}[tbp]
\centerline{\makebox{\includegraphics[width=3.5in]{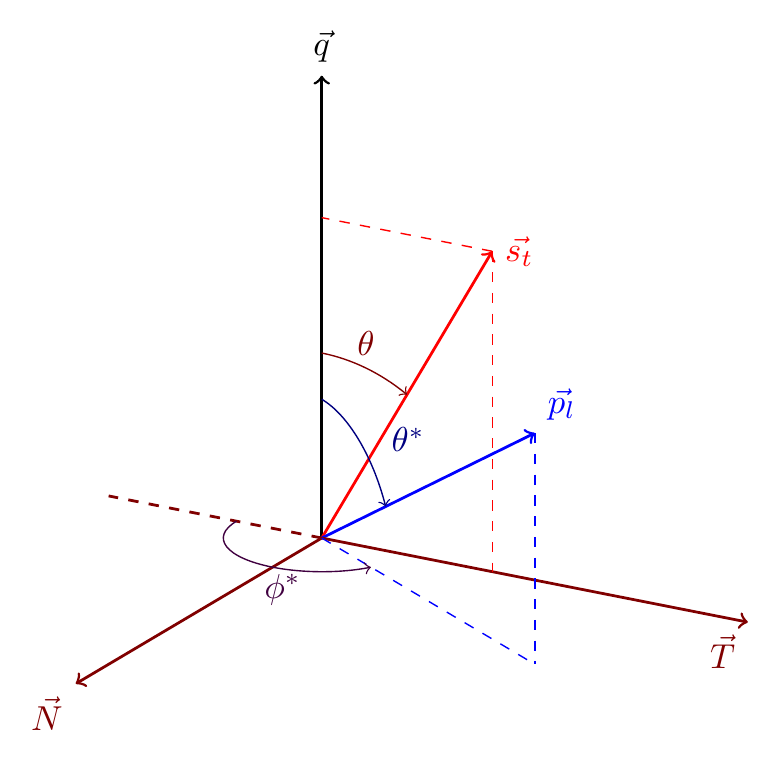}}} 
\caption{The coordinate system used for the analysis of polarized single top 
quarks produced in the $t$-channel.  A Cartesian coordinate system is defined
by the $\left\{ -{\vec T}, \vec{N}, {\vec q} \right \}$ directions. The angles
$\theta$, $\theta^*$ and $\phi^*$ are defined in the text.
\label{axes}}
\end{figure}

Among the six known quarks of the Standard Model (SM), the top quark is unique in
having a large mass, 173.2$\pm$0.9~GeV~\cite{ref:TeVAverage}, which is about 35 times as large as the next heavy quark, twice as large as the $W$ boson, and close the to the electroweak symmetry breaking scale $\Lambda_{EW}=1/(\sqrt{2}G_F)\approx$~246~GeV, where $G_F$ is the Fermi constant. The large mass of the top quark may
be a remnant of couplings between the top quark and new physics at a higher mass scale. The new physics can be described by an effective Lagrangian, and
one of its effects is to modify the structure of the $Wtb$ vertex. In the SM the $Wtb$ vertex is described by the coupling
\begin{displaymath}
{\cal L}_{Wtb} = -\frac{g}{\sqrt{2}}{\bar b}\gamma^\mu \frac{1}{2}\left
  (1-\gamma^5 \right)tW^-_\mu + \text{h.c.}
\end{displaymath}
Radiative corrections to the vertex can be absorbed into a small number of nonrenormalizable effective couplings called anomalous couplings, as follows:
\begin{displaymath}
{\cal L}_{Wtb} = -\frac{g}{\sqrt{2}}{\bar b}\gamma^\mu \left (V_LP_L+V_RP_R\right )tW^-_\mu  
-\frac{g}{\sqrt{2}}{\bar b}\frac{i\sigma^{\mu\nu}}{M_W}q_{\nu} \left (g_LP_L+g_RP_R\right)tW^-_\mu + \text{h.c.}
\end{displaymath}
In this expression, $P_L\equiv\frac{1}{2}(1-\gamma^5)$ and $P_R\equiv\frac{1}{2}(1+\gamma^5)$ 
are the left and right-handed projection operators, $V_L$, $V_R$, $g_L$, and $g_R$ are complex
coupling constants which all (except for $V_L$) vanish in the SM, and $q$ is
the four-momentum transfer at the $Wtb$ vertex. Imaginary components of any of
 coupling constants are particularly interesting since they signal new sources of $CP$ violation, beyond the single phase in the CKM matrix which is presently
the only known source. 
\begin{figure}[htbp]
\centerline{\makebox{\includegraphics[width=2.6in]{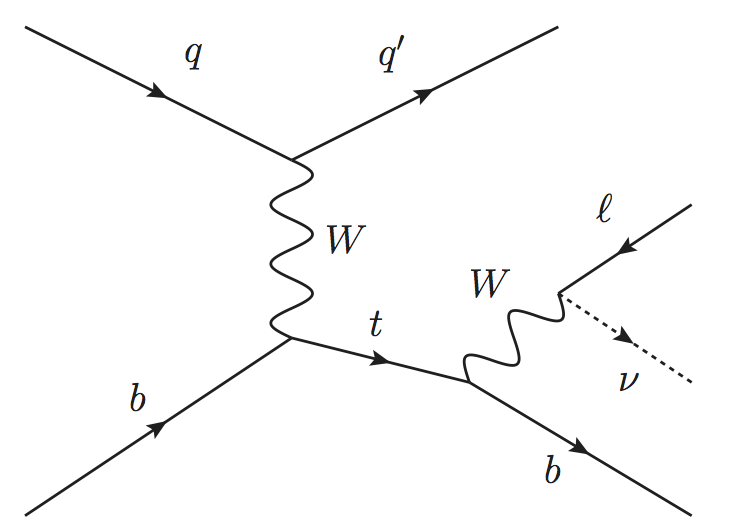}}}
\caption{$t$-channel production of single top quarks in $pp$ collisions.
\label{fig:tchannel}}
\end{figure}
The couplings influence the angular distribution of top quark decay products as described in ref.~\cite{AguilarSaavedra:2010wpol}. We shall be mainly interested in $t$-channel single quark production and decay at the LHC, illustrated in Fig.~\ref{fig:tchannel}.  Experimentally one can
detect and fully reconstruct the decay $t\rightarrow W^+b$; $W^+\rightarrow l^+\nu$ and measure the momentum of the spectator jet.  The neutrino in the decay
can be reconstructed from missing transverse energy in a ``hermetic'' detector
covering a solid angle of approximately 4$\pi$.  The coordinate system in which we describe the decay kinematics is as follows. The momentum of the $W$ boson in the top quark rest frame is called ${\vec q}$ and the spectator quark jet
direction, which we take as polarization axis of the top quark is called 
${\vec s}_t$.  Two  other directions  in space orthogonal to ${\vec q}$ are meaningful as proposed in ref.~\cite{AguilarSaavedra:2010wpol} and described
Fig.~\ref{axes}:
\begin{eqnarray}
{\vec N} & = & {\vec s}_t  \times {\vec q}\,,   \nonumber \\ 
{\vec T} & = & {\vec q}    \times {\vec N}\,. \nonumber 
\end{eqnarray}
Using these directions we can construct a right-handed coordinate system
such that the ${\hat x}$ direction points along $-{\vec T}$; the ${\hat y}$ direction lies along $\vec{N}$, and the ${\hat z}$ direction points along ${\vec q}$.
We define $\theta$ as the angle between ${\vec s}_t$ and ${\vec q}$ in the top quark rest frame.
The momentum of the charged lepton as measured in the $W$ rest frame is called $\vec{p_l}$.
In this paper we analyze the distribution of polar and azimuthal helicity angles, $\theta^*$ and $\phi^*$, of that lepton in this coordinate system.  
The main objective is to obtain normalized analytic expressions describing the probability densities in angular variables, parametrized in terms of 
observable physics quantities (to be defined below), which incorporate detector effects.  The purpose of these expressions is that they can then be used with
reconstructed collider data to build a likelihood function in a multidimensional
space of relevant physics parameters.  A variety of more or less well-known 
techniques use the likelihood function to draw inferences on the physics
parameters. These techniques include the unbinned maximum likelihood fit, 
the Feldman-Cousins method~\cite{feldmanCousins}, and Bayesian inference using Markov chain Monte 
Carlo. 

\section{Differential decay rates for polarized single top quark} 
\label{section:doubleDiff}
We construct the angular dependence of the decay $t\to W^+ b;W^+ \to l^+\nu$ using the helicity formalism~\cite{Richman:1984gh}.
The angular dependence of the amplitude for two-body decay is given by
\begin{equation}
A(a\to f) = \left[\frac{2J+1}{4\pi}\right]^{\frac{1}{2}} {D^J_{M,\lambda}}^*(\varphi,\theta,-\varphi)A_{\lambda_1,\lambda_2}\,,
\end{equation}
where $\lambda_1$ and $\lambda_2$ are the helicities of the outgoing particles and $\lambda=\lambda_1-\lambda_2$.  $J$ and $M$ are the spin and helicity of the decaying particle, $D^J_{M,L}$ is the Wigner D-function, and the angles are defined in the rest frame of the decaying particle.
$A_{\lambda_1,\lambda_2}$ is the amplitude for the decay to the specified helicity states.

One can construct the full triple differential for the angular distributions by including the amplitude for $t\to W^+ b$ and $W^+\to l^+\nu$.
For $W^+\to l^+{\nu}$, $J=1$, $\lambda_1=\frac{1}{2}$, $\lambda_2=-\frac{1}{2}$, and $\lambda=1$:
\begin{eqnarray}
A(t\to W^+ b) &=& \left[\frac{1}{2\pi}\right]^{\frac{1}{2}} {D^\frac{1}{2}_{\frac{1}{2},\lambda_W-\lambda_b}}^*(\varphi,\theta,-\varphi)A_{\lambda_W,\lambda_b}\,,\nonumber\\
A(W^+\to l^+\nu) &=& \left[\frac{3}{4\pi}\right]^{\frac{1}{2}} {D^1_{\lambda_W,1}}^*(\varphi^*,\theta^*,-\varphi^*)A_{\frac{1}{2},-\frac{1}{2}}\,.
\end{eqnarray}
We then obtain the total angular distributions from summing the amplitudes over the possible intermediate $W$ helicities, squaring the magnitude and summing over the possible final state $b$-helicities:
\begin{eqnarray}
\frac{dN}{d\Omega d\Omega^*} &=& \Sigma_{\lambda_b} \left| \Sigma_{\lambda_W} A(t\to W^+ b)A(W^+\to l^+\nu)\right|^2\\
&=& \Sigma_{\lambda_b} \left| \Sigma_{\lambda_W} \frac{\sqrt{6}}{4\pi} {D^\frac{1}{2}_{\frac{1}{2},\lambda_W-\lambda_b}}^*(\varphi,\theta,-\varphi)A_{\lambda_W,\lambda_b} {D^1_{\lambda_W,1}}^*(\varphi^*,\theta^*,-\varphi^*)A_{\frac{1}{2},-\frac{1}{2}}\right|^2. \nonumber
\end{eqnarray}
For $\lambda_b=\frac{1}{2}$, only intermediate $W$ helicities $\lambda_W=1,0$ are possible, while for $\lambda_b=-\frac{1}{2}$, $\lambda_W=-1,0$ are possible.
Thus the angular distribution contains 6 possible terms: 4 proportional to the square of each of the possible general amplitudes, and 2 for the two possible interference terms.
So although there are 4 general amplitudes with 3 relative phases, only 2 relative phases are accessible in the decay information for polarized single top quark decay.
The interference term between intermediate $W$s with helicity $\pm 1$ above has zero amplitude from the $t\to Wb$ decay.
Note that the direction defining $\varphi=0$ is arbitrary, but will be the same for $\varphi$ (the plane defined by the top quark polarization direction and the $W$ direction in the top quark rest frame), and $\varphi^*$ (the plane defined by the $W$ direction in the top quark rest frame and lepton direction in the $W$ rest frame).  Thus all dependencies appear in the combined formula as $\phi^*=\Delta\varphi = \varphi^*-\varphi$.

The total angular distribution for a spin up top quark is then
\begin{eqnarray}
\frac{1}{N}\frac{dN}{d\Omega d\Omega^*}&=&\frac{1}{(4\pi)^2} \left[\frac{3}{4}\left|A_{1,\frac{1}{2}}\right|^2\left(1+\cos\theta\right)\left(1+\cos\theta^*\right)^2 \right.\nonumber\\
 &+& \frac{3}{4}\left|A_{-1,-\frac{1}{2}}\right|^2 \left(1-\cos\theta\right)\left(1-\cos\theta^*\right)^2\nonumber\\
 &+& \frac{3}{2}\left(\left|A_{0,\frac{1}{2}}\right|^2(1-\cos\theta) + \left|A_{0,-\frac{1}{2}}\right|^2(1+\cos\theta)\right)\sin^2\theta^*\nonumber\\
 &-& \frac{3\sqrt{2}}{2} \sin\theta\sin\theta^*\left(1+\cos\theta^*\right)\Re\left({e^{i\phi^*}A_{1,\frac{1}{2}}A_{0,\frac{1}{2}}^*}\right)\nonumber\\
 &-& \left. \frac{3\sqrt{2}}{2} \sin\theta\sin\theta^*\left(1-\cos\theta^*\right)\Re\left({e^{-i\phi^*}A_{-1,-\frac{1}{2}}A_{0,-\frac{1}{2}}^*}\right)\right],
\label{eq:tripleDiffNoY}
\end{eqnarray}
where we have normalized the amplitudes, $A_{\lambda_W,\lambda_b}$ such that the total, $|\ARR|^2 + |\ALL|^2 + |\AZR|^2 + |\AZL|^2 =1$.
For reasons that will be clear in the following sections, we will prefer to express these
in terms of the spherical harmonics $Y_l^m(\theta, \phi)$; after including polarization,
\begin{multline}
  \varrho(\theta,\theta^*,\phi^*)=  \frac{1}{N} \frac{dN}{d\Omega d\Omega^{*}} \\
\begin{split}
    = \frac{1}{(4\pi)^{\frac{3}{2}}} \Biggl[\biggl\{ &\left(|\ARR|^2 + |\ALL|^2 + |\AZR|^2 + |\AZL|^2\right) \Pt{0}{0} \\
    + P &\left(|\ARR|^2 - |\AZR|^2 + |\AZL|^2 - |\ALL|^2 \right) \Pt{1}{0} \biggr\} \Ys{0}{0} \\
    + \frac{\sqrt{3}}{2} \biggl\{&\left( |\ARR|^2 - |\ALL|^2 \right) \Pt{0}{0} \\
    + P &\left( |\ARR|^2 + |\ALL|^2 \right) \Pt{1}{0} \biggr\} \Ys{1}{0} \\
    + \frac{1}{2\sqrt{5}} \biggl\{ &\left(|\ARR|^2 - 2|\AZR|^2 - 2|\AZL|^2 + |\ALL|^2\right) \Pt{0}{0} \\
    + P &\left(|\ARR|^2 + 2|\AZR|^2 - 2|\AZL|^2 - |\ALL|^2\right) \Pt{1}{0} \biggr\} \Ys{2}{0} \\
    - P \biggl\{ \frac{\sqrt{3}}{2} &\left( \ARR \AZR^{*} + \ALL^{*} \AZL \right) \Pt{1}{1} \Ys{1}{1} \\
    - \frac{\sqrt{3}}{2} &\left( \ARR^{*} \AZR + \ALL \AZL^{*} \right) \Pt{1}{1} \Ys{1}{-1} \\
    +\frac{\sqrt{3}}{2\sqrt{5}} &\left( \ARR \AZR^{*} - \ALL^{*} \AZL \right) \Pt{1}{1} \Ys{2}{1}\\
    - \frac{\sqrt{3}}{2\sqrt{5}} &\left( \ARR^{*} \AZR - \ALL \AZL^{*} \right) \Pt{1}{1} \Ys{2}{-1} \biggr\} \Biggr].
\end{split}
    \label{eq:trippleDiff}
\end{multline}
By integrating over the $t\to W^+ b$ angles, we get the angular distribution for the $W$-decay from polarized top quark,
\begin{equation}
\begin{split}
\frac{1}{N}\frac{dN}{d\Omega^*}=\frac{1}{\sqrt{4\pi}} \Biggl[&\left(\left|A_{1,\frac{1}{2}}\right|^2 + \left|A_{-1,-\frac{1}{2}}\right|^2  + \left|A_{0,\frac{1}{2}}\right|^2 + \left|A_{0,-\frac{1}{2}}\right|^2 \right)Y_0^0\left(\theta^*, \phi^*\right) \\
 + \frac{\sqrt{3}}{2}&\left(\left|A_{1,\frac{1}{2}}\right|^2 - \left|A_{-1,-\frac{1}{2}}\right|^2\right)Y^0_1(\theta^*,\phi^*)\\
 + \frac{1}{\sqrt{5}}&\left(\frac{1}{2}\left(\left|A_{1,\frac{1}{2}}\right|^2 + \left|A_{-1,-\frac{1}{2}}\right|^2\right)-\left(\left|A_{0,\frac{1}{2}}\right|^2 + \left|A_{0,-\frac{1}{2}}\right|^2\right)\right)Y^0_2(\theta^*,\phi^*) \\
 + P\biggl\{\frac{\sqrt{3}\pi}{8}&\left(A_{1,\frac{1}{2}}A_{0,\frac{1}{2}}^*+A_{-1,-\frac{1}{2}}^*A_{0,-\frac{1}{2}}\right)Y^1_1(\theta^*,\phi^*) \\
 - \frac{\sqrt{3}\pi}{8}&\left(A_{1,\frac{1}{2}}^*A_{0,\frac{1}{2}}+A_{-1,-\frac{1}{2}}A_{0,-\frac{1}{2}}^*\right)Y^{-1}_1(\theta^*,\phi^*) \\
 + \frac{\sqrt{3}\pi}{8\sqrt{5}}&\left(A_{1,\frac{1}{2}}A_{0,\frac{1}{2}}^*-A_{-1,-\frac{1}{2}}^*A_{0,-\frac{1}{2}}\right)Y^1_2(\theta^*,\phi^*) \\
 - \frac{\sqrt{3}\pi}{8\sqrt{5}}&\left(A_{1,\frac{1}{2}}^*A_{0,\frac{1}{2}}-A_{-1,-\frac{1}{2}}A_{0,-\frac{1}{2}}^*\right)Y^{-1}_2(\theta^*,\phi^*)\biggr\}\Biggr].
\end{split}\label{eq:doubleDiff1}
\end{equation}
A purely real form of this expression after translating the $A_{\lambda_W,\lambda_b}$ to the eight form-factors of ref.~\cite{AguilarSaavedra:2010wpol} is
\begin{eqnarray}
\frac{dN}{d\Omega^*}&=&\frac{1}{(4\pi)} \left[A_0 + 2 B_0\right.\nonumber\\
 &+& \left(6\frac{|\vec{q}|}{m_t}B_1\right)P^0_1(\cos\theta^*)\nonumber\\
 &+& \left(B_0-A_0\right)P^0_2(\cos\theta^*)\nonumber \\
 &+& P\frac{3\pi}{4} \left(\frac{m_t}{M_W}C_0\cos\phi^* - \frac{|\vec{q}|}{M_W}D_1\sin\phi^*\right)P^1_1(\cos\theta^*)\nonumber\\
 &+& P\left. \frac{\pi}{4} \left(\frac{|\vec{q}|}{M_W}C_1\cos\phi^* - \frac{m_t}{M_W}D_0\sin\phi^*\right)P^1_2(\cos\theta^*)\right].
\label{eq:doubleDiff2}
\end{eqnarray}
One more integration of eq.~\ref{eq:doubleDiff1} over $\phi^*$ gives a decay rates in $\theta^{*}$ alone:
\begin{eqnarray}
\frac{1}{N}\frac{dN}{d\theta^*}&=&\sqrt{\pi} \left[\left(\left|A_{1,\frac{1}{2}}\right|^2 + \left|A_{-1,-\frac{1}{2}}\right|^2  + \left|A_{0,\frac{1}{2}}\right|^2 + \left|A_{0,-\frac{1}{2}}\right|^2 \right)Y_0^0\left(\theta^*, \phi^*\right) \right.\nonumber\\
 &+& \sqrt{3}\pi\left(\left|A_{1,\frac{1}{2}}\right|^2 - \left|A_{-1,-\frac{1}{2}}\right|^2\right)Y^0_1(\theta^*,\phi^*)\nonumber\\
 &+& \frac{2\pi}{\sqrt{5}}\left(\frac{1}{2}\left(\left|A_{1,\frac{1}{2}}\right|^2 + \left|A_{-1,-\frac{1}{2}}\right|^2\right)-\left(\left|A_{0,\frac{1}{2}}\right|^2 + \left|A_{0,-\frac{1}{2}}\right|^2\right)\right)Y^0_2(\theta^*,\phi^*)\,.
\label{eq:doubleDiffStar}
\end{eqnarray}

\section {Angular decomposition and parametrization}
The double differential decay rate,  eq.~\ref{eq:doubleDiff1}, is expressed in spherical harmonics
because it simplifies the incorporation of detector effects.  The simplifications result from the 
ortho-normality of the spherical harmonics on one hand, and a ``spherical'' version of the convolution
theorem on the other hand. Before proceeding to deal with these we first obtain a normalized 
joint probability density 
\begin{equation}
\rho(\theta^*, \phi^*; \vec {\alpha})\,,
\end{equation}
the ${\vec \alpha}$ referring to the parameters which are to be estimated
using likelihood techniques.   Given a dataset with $n$ events ${\cal D}=\left\{ (\theta^*_0, \phi^*_0 ), (\theta^*_1, \phi^*_1)... (\theta^*_{n-1}, \phi^*_{n-1}) \right\}$, one constitutes
a likelihood function 
\begin{equation}
{\cal L}({\vec \alpha}) = \rho({\cal D}; {\vec \alpha}) = \prod_{i=0}^{n-1} \rho(\theta^*_i, \phi^*_i; \vec {\alpha})\,.
\label{eq:lik1}
\end{equation}
For a sufficiently well-behaved likelihood function, the simplest technique, 
an unbinned maximum likelihood fit, can be applied to obtain what is known
as the maximum likelihood estimator (MLE), based on the quantity 
\begin{equation}
-2\ln{\cal L}({\vec \alpha})  = -2\sum_{i=0}^{n-1} \ln {\rho(\theta^*_i, \phi^*_i; \vec {\alpha})}\,.
\label{eq:lik2}
\end{equation}
With large data samples $-2\ln{\cal L}({\vec \alpha})$ is often parabolic (and ${\cal L}({\vec \alpha})$ Gaussian), and the technique is applicable. 
The central value occurs at $\vec{\alpha}_{0}$, the point in parameter space for which $-2\ln{\cal {\cal L\mathrm{(}\vec{\alpha}}\mathrm{)}}$ 
is minimized over the parameters $\vec{\alpha}$. The covariance matrix for the parameter $\vec{\alpha}$ is $\mathbf{C}=\left(\mathbf{H}/2\right)^{-1}$, 
where $\mathbf{H}$ is the Hessian matrix of $-2\ln{\cal L}({\vec \alpha})$ near $\vec{\alpha}_{0}$. More complicated techniques, beyond the scope of this
paper, are generally required when the likelihood function is non-Gaussian. However, the log-likelihood, eq.~\ref{eq:lik1}, is still an important ingredient. 

We require a set of parameters which always gives a positive-definite probability density function (PDF) over the entire range of allowed values.  Existing numerical minimization 
code~\cite{ref:MINUIT} is easier to work with if the physically allowed region of the parameter space is rectangular. Such a parametrization is 
\begin{eqnarray}
f_{1} & = &\frac {\left(\left|A_{1,\frac{1}{2}}\right|^2 + \left|A_{-1,-\frac{1}{2}}\right|^2 \right)} {\left(\left|A_{1,\frac{1}{2}}\right|^2 + \left|A_{-1,-\frac{1}{2}}\right|^2  + \left|A_{0,\frac{1}{2}}\right|^2 + \left|A_{0,-\frac{1}{2}}\right|^2 \right)} \nonumber \\
f_{1}^+ & = & \frac {\left(\left|A_{1,\frac{1}{2}}\right|^2\right)} {\left(\left|A_{1,\frac{1}{2}}\right|^2 + \left|A_{-1,-\frac{1}{2}}\right|^2 \right)} \nonumber \\
f_{0}^+ & = & \frac {\left(\left|A_{0,\frac{1}{2}}\right|^2\right)} {\left(\left|A_{0,\frac{1}{2}}\right|^2 + \left|A_{0,-\frac{1}{2}}\right|^2 \right)} \nonumber \\
\delta_+ & = & {\rm arg} \left(A_{1,\frac{1}{2}}A_{0,\frac{1}{2}}^* \right) \nonumber \\
\delta_- & = & {\rm arg} \left(A_{-1,-\frac{1}{2}}A_{0,-\frac{1}{2}}^* \right) 
\label{eq:paramDef}
\end{eqnarray}
where $f_1$, $f_1^+$, $f_0^+\in [0,1]$.  The space of observable parameters is five-dimensional. This counting does not include the production cross section since the
PDF is normalized and it does not include the polarization.  From ref.~\cite{ref:minimalSet}, one would be expect a \emph{six} dimensional parameter space; the explanation 
for this reduction is the lack of interference between the $t\rightarrow l^+ \nu b_r$ and $t \rightarrow l^+ \nu b_l$ transitions, which we have already noted. 
Thus while the parametrization of Ref~\cite{ref:minimalSet} is ``minimal'' in a theoretical sense, that of eq.~\ref{eq:paramDef} is minimal in an experimental sense 
and is sufficient to describe all observable phenomena in the decay of $t\rightarrow W^+b\rightarrow l^+ \nu b$. 

We will use the shorthand $\vec{\alpha}$ for the set of parameters $\vec{\alpha}\equiv\left \{ f_1, f_1^+, f_0^+, \delta_+, \delta_-, P\right\}$. Eq.~\ref{eq:doubleDiff1} is normalized
\begin{equation}
  \int\rho(\theta^*, \phi^*; \vec {\alpha}) d\Omega = 1
\end{equation}
as required by likelihood methods.  We rewrite this equation as
\begin{equation}
  \rho(\theta^*, \phi^*; \vec {\alpha}) =  \frac{1}{N} \frac{dN}{d\Omega} = \sum_{l=0}^2 \sum_{m=-l}^l a_{l,m}(\vec{\alpha}) Y_l^m(\theta^*,\phi^*)\,.
\end{equation}
where
\begin{eqnarray}
a_{0,0} & = & \frac {1}{\sqrt{4\pi}}\,,                           \nonumber \\
a_{1,0} & = & \frac {\sqrt{3}}{\sqrt{4\pi}} f_1 (f_1^+ - \frac {1}{2}) \,,\nonumber \\
a_{2,0} & = & \frac {1}{\sqrt{20\pi}}(\frac{3}{2}f_1 - 1) \,,      \nonumber \\
a_{1,1}  & = & +P \cdot \frac{\sqrt{3\pi}}{16} \sqrt{f_1(1-f_1)} \left \{  \sqrt{f_1^+f_0^+}e^{i\delta_+} +  \sqrt{(1-f_1^+)(1-f_0^+)}e^{-i\delta_-}     \right \}\,,\nonumber \\
a_{1,-1} & = & -P \cdot \frac{\sqrt{3\pi}}{16} \sqrt{f_1(1-f_1)} \left \{  \sqrt{f_1^+f_0^+}e^{-i\delta_+} +  \sqrt{(1-f_1^+)(1-f_0^+)}e^{i\delta_-}     \right \}\,,\nonumber \\
a_{2,1}  & = & +P \cdot \frac{\sqrt{3\pi}}{16\sqrt{5}} \sqrt{f_1(1-f_1)} \left \{  \sqrt{f_1^+f_0^+}e^{i\delta_+} -  \sqrt{(1-f_1^+)(1-f_0^+)}e^{-i\delta_-}     \right \}\,,\nonumber \\
a_{2,-1} & = & -P \cdot \frac{\sqrt{3\pi}}{16\sqrt{5}} \sqrt{f_1(1-f_1)} \left \{  \sqrt{f_1^+f_0^+}e^{-i\delta_+} -  \sqrt{(1-f_1^+)(1-f_0^+)}e^{i\delta_-}     \right \}\,,
\label{eq:physicsCoefficients}
\end{eqnarray}
and all of the other coefficients are zero. The coefficients in this angular decomposition are 
an alternate representation of the PDF; they contain the same information as in 
eq.~\ref{eq:doubleDiff1}. 
The likelihood function is defined by the dataset and depends on the parameters  $\vec{\alpha}_{0}$ through the coefficients. For each point $(\theta^*_i, \phi^*_i)\in \cal{D}$ define
\begin{equation}
y^i_{l,m} = Y_{l,m}(\theta^*_i, \phi^*_i)\,.
\end{equation}
Our likelihood is 
\begin{equation}
-2\ln{\cal L}({\vec \alpha})  = -2\sum_{i=0}^{n-1} \ln {\sum_{l=0}^2 \sum_{m=-l}^l a_{l,m}(\vec{\alpha}) y^i_{l,m}} \,.
\label{eq:indexmath1}
\end{equation}
Henceforth, for convenience in this note we will always imply summation over repeated indices; summation
over $l$ indices is from zero through an appropriate upper limit (implied or explicitly stated), while
summation over $m$ indices is from $-l$ to $l$.  Eq.~\ref{eq:indexmath1} can be written as 
\begin{equation}
-2\ln{\cal L}({\vec \alpha})  = -2\sum_{i=0}^{n-1} \ln{\left(a_{l,m}(\vec{\alpha} y^i_{l,m})\right)}\,.
\label{eq:indexmath2}
\end{equation}

\section {Detector effects}

In this section we incorporate detector effects, specifically efficiency, resolution, and background.  Efficiency includes trigger efficiency, 
selection efficiency, and acceptance.  Angular resolution receives contributions from both the reconstruction of the lepton momentum and the 
reconstruction of the coordinate axes in Fig.~\ref{axes}; these will also be affected by the procedures used to compute the neutrino momentum 
from the lepton momentum and the missing transverse energy.  Physics effects such as gluon radiation and fragmentation can be absorbed into 
efficiency and/or resolution.  Whenever the radiation of a hard gluon causes an event to be mismeasured, it contributes to the resolution; when 
it causes the event to fail acceptance cuts it contributes to efficiency.  Single top quark backgrounds typically include $t{\bar t}$ production, 
$W$+jet production, and multijet production. 

\begin{figure}[htbp]
\centerline{\makebox{
    \includegraphics[width=4.5in]{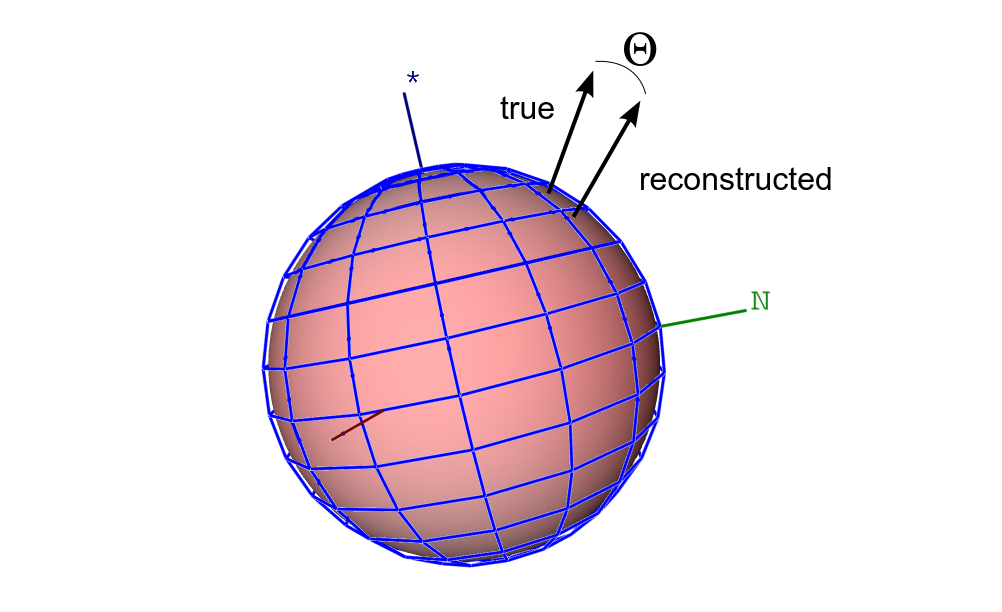}}}
    \caption{The angular deviation is the deviation $\Theta$  between the reconstructed direction of the lepton, in the canonical 
             coordinate system defined by the directions $\left\{ {-\vec T}, {\vec N}, {\vec q} \right\}$, and the true direction. 
             \label{fig:angleIllustrate}}
\end{figure}

Suppose that empirical descriptions of the detector efficiency, angular resolution, and background are available, and furthermore that these descriptions 
are also expressed in a spherical decomposition, along the same lines as that just obtained for the physics PDF.  These could, for example, be obtained 
from a fit to Monte Carlo.  Specifically, we assume that:
\begin{itemize}
\item {the efficiency is parametrized as $\epsilon(\theta^*, \phi^*) = e_{l,m}Y_l^m(\theta^*,\phi^*)$;}
\item {the background is parametrized as $\beta(\theta^*, \phi^*) = b_{l,m}Y_l^m(\theta^*,\phi^*)$;}
\item {the angular resolution is parametrized as ${\cal R} (\Theta) = r_l P_l(\cos{\Theta})$, where $\Theta$
  is the angular deviation between measured and true lepton directions (see Fig.~\ref{fig:angleIllustrate}), and}
\item {the signal constitutes a fraction $f_s$ of the data sample and the background constitutes a fraction $1-f_s$. }
\end{itemize}
All of the above expansions (efficiency, resolution, background) can be constructed to be real and positive-definite. The
efficiency expansion is assumed to be less than or equal to unity for all value of $\theta^*$ and $\phi^*$, and the resolution
and background descriptions are normalized.  The two angles $\theta^*$ and $\phi^*$ are polar and azimuthal angles of the decay
lepton in the basis defined by the directions $\left\{ {-\vec T}, {\vec N}, {\vec q} \right\}$.  

As we will prove below, \emph {all of the above detector 
effects are incorporated analytically} by replacing the coefficients 
$a_{l,m}(\vec{\alpha})$ by new coefficients ${\cal A}_{l,m}(\vec{\alpha})$ according to
\begin{equation}
a_{l,m}(\vec{\alpha})\rightarrow{\cal A}_{l,m}(\vec{\alpha}) \equiv
f_s{\frac {2}{2l+1}}\frac{G_{\lambda,\lambda^\prime,l}^{\mu,\mu^\prime,m} a_{\lambda,\mu}  (\vec{\alpha}) e_{\lambda^\prime,\mu^\prime } r_{l}}{(-1)^{\sigma}  a_{\kappa,\sigma}(\vec{\alpha}) {e_{\kappa,-\sigma} }} 
+ (1-f_s)b_{l,m}\,.
\label{eq:coeff}
\end{equation}
In this expression $G_{\lambda,\lambda^\prime,l}^{\mu,\mu^\prime,m}$ is the Gaunt coefficient,  related to the
Clebsh-Gordan coefficients $C^{\mu,\mu^\prime,m}_{\lambda,\lambda^\prime,l}$ by the equation:
\begin{equation}
G_{\lambda,\lambda^\prime,l}^{\mu,\mu^\prime,m}=\sqrt{\frac{(2\lambda+1)(2\lambda^\prime + 1)}{4\pi (2l+1)}} C^{0,0,0}_{\lambda \lambda^\prime l} C^{\mu,\mu^\prime,m}_{\lambda,\lambda^\prime,l}       \label{eq:GauntDef}\,.
\end{equation}

\noindent {\bf Proof} Start with the ``raw physics'' coefficients $a_{l,m}$ given in eq.~\ref{eq:physicsCoefficients}. 
First we will add detector efficiency effects, then we will add resolution effects, and finally
we will add the background.  Denote the PDF before these effects as
\begin{equation}
\rho_p(\theta^*, \phi^*; \vec{\alpha}) = a_{l,m}(\vec{\alpha}) Y_l^m(\theta^*, \phi^*).
\end{equation}
To include detector efficiency, we multiply by the efficiency function $\epsilon(\theta^*, \phi^*; \vec{\alpha})$ and normalize
again
\begin{eqnarray}
  \rho_e(\theta^*, \phi^*; \vec{\alpha}) &=& N \cdot \epsilon(\theta^*, \phi^*; \vec{\alpha}) \rho_p(\theta^*, \phi^*; \vec{\alpha}) \nonumber \\
                                     &=& N \cdot e_{l^\prime,m^\prime}Y_{l^\prime,m^\prime}(\theta^*, \phi^*) a_{l,m}(\vec{\alpha}) Y_l^m(\theta^*, \phi^*)
\end{eqnarray} 
where $N$ is a normalization constant defined as
\begin{eqnarray}
  1/N & = &  \int_{4\pi} \rho_p(\theta^*, \phi^*) \epsilon(\theta^*,\phi^*)   d\Omega \nonumber \\
  & = &   \int_{4\pi} e_{l^\prime,m^\prime} Y_{l^\prime}^{m^\prime} (\theta^*, \phi^*) a_{l,m}(\vec{\alpha}) Y_l^m (\theta^*, \phi^*)    d\Omega \nonumber  \\
  & = &   (-1)^{m^\prime} a_{l,m}(\vec{\alpha}) e_{l^\prime,m^\prime} \int_{4\pi}  Y_l^m (\theta^*, \phi^*)Y_{l^\prime}^{-m^\prime *} (\theta^*, \phi^*) d \Omega \nonumber \\
  & = &    (-1)^{m^\prime} a_{l,m}(\vec{\alpha}) e_{l^\prime,m^\prime} \delta_{l,l^\prime} \delta_{m,-m^\prime}\nonumber\\
  & = &    (-1)^{m} a_{l,m}(\vec{\alpha}) e_{l,-m} 
  \label{eq:normFactor}
\end{eqnarray}
so that
\begin{eqnarray}
  \rho_e(\theta^*, \phi^*; \vec{\alpha}) & = & N\cdot  \epsilon(\theta^*, \phi^*)  \rho_e(\theta^*, \phi^*; \vec{\alpha})\nonumber\\
& = &\frac{e_{l^\prime,m^\prime}Y_{l^\prime,m^\prime}(\theta^*, \phi^*) a_{l,m}(\vec{\alpha}) Y_l^m(\theta^*, \phi^*)}  {(-1)^{m} a_{l,m}(\vec{\alpha}) e_{l,-m} } \nonumber \\
                     & = &  \frac{ a_{l,m} (\vec{\alpha}) Y_l^m (\theta^*, \phi^*) e_{l^\prime, m^\prime } Y_{l^\prime}^{m^\prime} (\theta^*, \phi^*)}  {(-1)^{\sigma} a_{\kappa,\sigma}(\vec{\alpha}) e_{\kappa,-\sigma} }\nonumber \\
                     & = &  \frac{G_{l, l^\prime, L}^{m, m^\prime, M} a_{l,m}  (\vec{\alpha}) e_{l^\prime, m^\prime } Y_L^M (\theta^*, \phi^*)}  {(-1)^{\sigma} a_{\kappa,\sigma}(\vec{\alpha}) e_{\kappa,-\sigma} }\nonumber \\
                     & = &  c_{L,M} (\vec{\alpha}) Y_L^M (\theta^*, \phi^*)\,,
\end{eqnarray} 
where the coefficients $c_{l,m}$ are
\begin{equation}
c_{l,m} (\vec{\alpha})= \frac{G_{\lambda, \lambda^\prime, l}^{\mu, \mu^\prime, m} a_{\lambda,\mu}  (\vec{\alpha}) e_{\lambda^\prime, \mu^\prime }}{(-1)^{\sigma} a_{\kappa,\sigma}(\vec{\alpha}) {e_{\kappa,-\sigma} }}\,.
\label{eq:cCoefficients}
\end{equation}
Here we have used Gaunt's theorem\cite{ref:Gaunt} for the product of two spherical harmonics, 
\begin{equation}
Y_\lambda^\mu(\theta^*, \phi^*)Y_{\lambda^\prime}^{\mu^\prime}(\theta^*, \phi^*) = G_{\lambda, \lambda^\prime, l}^{\mu, \mu^\prime, m}Y_l^m(\theta^*, \phi^*)\,.
\end{equation}
Thus far we have a normalized PDF for the signal including detector efficiency which is in the same form as the
original PDF and represented by the coefficients $c_{l,m}$ given by eq.~\ref{eq:cCoefficients}. We are now going to 
add the detector smearing to that by convolving the signal PDF
\begin{equation}
  \rho_e(\theta^*, \phi^*; \vec{\alpha})= c_{l,m}(\vec{\alpha})Y_l^m(\theta^*, \phi^*)
\end{equation}
with the convolution kernel, a linear combination of $M+1$ Legendre polynomials:
\begin{equation}
  {\cal R} (\Theta) = \sum_{l=0}^M r_l P_l(\cos{\Theta}).
\end{equation}
To do this we use an analogue of the convolution theorem.   Imagine that $f$ and $g$ are 
simple functions of one variable. The Fourier transform of the convolution $f*g$ is the product of 
the Fourier transforms of $f$ and $g$. In spherical polar coordinates the analogous theorem is the Funk-Hecke
theorem. It is ideally suited to our needs, because we have been dealing all along with coefficients
of an expansion in spherical harmonics.  The theorem states~\cite{ref:funkHecke,ref:Groemer} that
\begin{eqnarray}
  \rho_e (\theta^*, \phi^*; \vec{\alpha}) * {\cal R}(\cos{\Theta}) &=&  \sum_{n}^M \sum_{m=-n}^n \frac {2}{2n+1} r_{n}c_{n,m}(\vec{\alpha}) Y_n^m(\theta^*, \phi^*) \nonumber \\
  &=&  d_{l,m}(\vec{\alpha})Y_l^m(\theta^*, \phi^*) 
\label{eq:fh1}
\end{eqnarray}
where 
\begin{eqnarray}
d_{l,m} (\vec{\alpha}) & \equiv & {\frac {2}{2l+1}} r_{l} c_{l,m} (\vec{\alpha}) \nonumber \\
       &   =    & {\frac {2}{2l+1}}\frac{G_{\lambda, \lambda^\prime, l}^{\mu, \mu^\prime, m} a_{\lambda\mu}  (\vec{\alpha}) e_{\lambda^\prime \mu^\prime } r_{l}}{(-1)^{\sigma}  a_{\kappa,\sigma}(\vec{\alpha}) {e_{\kappa,-\sigma} }} 
\label{eq:fh2}
\end{eqnarray}
(no summation over l) which vanish for $l>M$.  

Finally, we add the background by taking a simple linear combination of the
the coefficients.  The coefficients for the full PDF including efficiency, resolution, and background are therefore:
\begin{equation}
{\cal A}_{l,m}(\vec{\alpha})=f_s{\frac {2}{2l+1}}\frac{G_{\lambda, \lambda^\prime, l}^{\mu, \mu^\prime, m} a_{\lambda,\mu}  (\vec{\alpha}) e_{\lambda^\prime, \mu^\prime } r_{l}}{(-1)^{\sigma}  a_{\kappa,\sigma}(\vec{\alpha}) {e_{\kappa,-\sigma} }} 
+ (1-f_s)b_{l,m}\,.
\label{eq:coeffFormula}
\end{equation}
There is no summation over $l$ in this formula. 
\subsection {Background}

In this section we outline a method to obtain the expansion coefficients $b_{l,m}$ of an angular decomposition of the background function $\beta(\theta^*, \phi^*)$. 
The input to the method is a dataset ${\cal B}=\left\{ (\theta^*_0, \phi^*_0 ), (\theta^*_1, \phi^*_1)... (\theta^*_{n-1}, \phi^*_{n-1}) \right\}$ of synthetic data 
from simulation, real data from a control sample, or a mix. For example, it could consist of simulated background events which have been subject to the same 
reconstruction procedures as actual collider data.  The background coefficients can be obtained by fitting this dataset.  Two steps are required.

First, a parametrized, real, positive-definite background function is constructed:
\begin{equation}
\beta(\theta^*, \phi^*) = \left |\sum_{l,m} g_{l,m}Y_l^m(\theta^*, \phi^*)\right|^2      
\label{eq:bkgDeff}
\end{equation}
where the $g_{l,m}$ are coefficients satisfying 
\begin{equation}
\sum_{l,m} |g_{l,m}|^2 = 1\,.
\label{eq:normDef}
\end{equation}
The constants $g_{l,m}$ can be determined in a likelihood fit to the dataset  ${\cal B}$. Then, the $b_{l,m}$ are obtained from the $g_{l,m}$ in the following way.  First, write 
\begin{eqnarray}
\beta(\theta^*, \phi^*) &=& g_{l,m}g^*_{l^\prime,m^\prime}Y_l^m(\theta^*, \phi^*)Y_{l^\prime}^{m^\prime *}(\theta^*, \phi^*)    \nonumber \\
&=& g_{l,m}g^*_{l^\prime,m^\prime}(-1)^{m^\prime} Y_l^m(\theta^*, \phi^*)Y_{l^\prime}^{-m^\prime }(\theta^*, \phi^*)    \nonumber \\
& = &  g_{l,m}g^*_{l^\prime,m^\prime}(-1)^{m^\prime} G_{l,l^\prime,L}^{m,-m^\prime,M}Y_L^M(\theta^*, \phi^*) 
\end{eqnarray}
so that the translation from the  $g_{l,m}$ to the $b_{l,m}$ is via the relation:
\begin{equation}
b_{L,M} =  g_{l,m}g^*_{l^\prime,m^\prime}(-1)^{m^\prime}\sqrt{\frac{(2l+1)(2l^\prime + 1)}{4\pi (2L+1)}} C^{0,0,0}_{l,l^\prime,L} C^{m,-m^\prime,M}_{l,l^\prime,L}
\label{eq:coeffTranslate}
\end{equation}
where we have expressed the Gaunt coefficients in terms of the Clebsh-Gordan coefficients using eq.~\ref{eq:GauntDef}. \emph{To summarize}, we first devise a function 
which is by construction a real-valued positive definite function to describe the background. We then fit that function to the background dataset ${\cal B}$ to obtain 
the background coefficients, which can then be translated into a more convenient form using eq.~\ref{eq:coeffTranslate}. The fit for $g_{l,m}$ is based on a likelihood function 
\begin{equation}
{\cal L}(g_{l,m}) = \beta({\cal B}; {g_{l,m}}) = \prod_{i=0}^{n-1} \beta(\theta^*_i, \phi^*_i; g_{l,m})\,.
\end{equation}
The quantity
\begin{eqnarray}
-2\ln{\cal L}(g_{l,m})  & = & -2\sum_{i=0}^{n-1} \ln {\beta(\theta^*_i, \phi^*_i; g_{l,m})} \nonumber \\
                       & = & -2\sum_{i=0}^{n-1} \ln |g_{l,m} \cdot y^i_{l,m}|^2\,,
\end{eqnarray}
where $y^i_{l,m}\equiv Y_l^m(\theta^*_i,\phi^*_i)$, is computed and minimized over all of the coefficients $g_{l,m}$.  When weighted events are available, the weight $w_i$ can be incorporated as a Bayesian prior in the likelihood function, which leads to only a slight modification of the above formula:
\begin{equation}
-2\ln{\cal L}(g_{l,m}) =  -2\sum_{i=0}^{n-1} w_i \ln |g_{l,m} \cdot y^i_{l,m}|^2\,.
\label{eq:backgroundMaster}
\end{equation}

The fits are best carried out using a parametrization of the coefficients in eq.~\ref{eq:normDef} which insures a positive-definite PDF, within rectangular bounds 
on the parameters. Such a parametrization prevents the fit from straying into physically disallowed portions of the parameter space, where the usual punishment 
to the user is a negative likelihood function with an undefined logarithm, followed by a core dump during the minimization step.

The phase of the $l=0$ component will be taken to be zero. For each additional set of components $l$, $m=\{-l, -l+1,... l-1, l\}$ we add $2(2l+1)$ additional parameters 
defined in the following way:
\begin{itemize}
\item {one fraction 
\begin{displaymath}
F_l\equiv\sum_{i \ge l}^{l_{max}} \sum_{m=-i}^i|g_{i,m}|^2 ;
\end{displaymath}
}
\item {one phase $\varphi_{l,0}\equiv \arg{(g_{l,0})}$}; 
\item {then, for each $m\in 1,2,...l$ }
\begin{itemize}
\item {one fraction 
\begin{displaymath}
F_{l,m}\equiv \sum_{i \ge m}^l(|g_{l,i}|^2 + |g_{l,-i}|^2);
\end{displaymath}
} 
\item {one fraction 
\begin{displaymath}
s_{l,m}\equiv \frac{|g_{l,m}|^2}{|g_{l,m}|^2 + |g_{l,-m}|^2};
\end{displaymath}
}
\item {one phase $\varphi_{l,m}\equiv\arg{(g_{l,m})}$}; 
\item {one phase $\varphi_{l,-m}\equiv\arg{(g_{l,-m})}$}. 
\end{itemize}
\end{itemize}
All of the fractions can vary on the interval [0, 1], while the phases need not be bounded at all. In the appendix we show that this minimization can be 
carried out \emph{analytically} for a very fast estimate of the background shape parameters.  From the estimated values of the $g_{l,m}$ we obtain the 
estimated value of the $b_{l,m}$ with the use of eq.~\ref{eq:coeffTranslate}.

\subsection {Efficiency}
\label{sec:efficiency}
In this section we outline a method to obtain the expansion coefficients $e_{l,m}$ of 
an angular decomposition of the efficiency function $\epsilon(\theta^*, \phi^*)$.  The input to the method is simulated signal events which have been subject to the same reconstruction procedures as actual collider data. The dataset contains not only those events which pass the selection criteria, but also those which fail. The coefficients
can be obtained by fitting the simulated data as we outline here.  Two steps are required.

First, a parametrized, real, positive-definite efficiency function is constructed:
\begin{equation}
\epsilon(\theta^*, \phi^*) = A^2\cdot \left |\sum_{l,m} f_{l,m}Y_l^m(\theta^*, \phi^*)\right|^2      
\label{eq:effDeff}
\end{equation}
where the $f_{l,m}$ are coefficients satisfying 
\begin{equation}
\sum_{l,m} |f_{l,m}|^2 = 1
\label{eq:effNormDef}
\end{equation}
and $A$ is an overall efficiency factor. The constants $f_{l,m}$  and the overall efficiency factor $A$ can be determined in a likelihood fit to signal Monte Carlo. Then, the $e_{l,m}$ are obtained from the $f_{l,m}$ through a set of algebraic manipulations that parallel those of eq.~\ref{eq:coeffTranslate} in order to obtain:
\begin{equation}
e_{L,M} =  A^2 f_{l,m}f^*_{l^\prime,m^\prime}(-1)^{m^\prime}\sqrt{\frac{(2l+1)(2l^\prime + 1)}{4\pi (2L+1)}} C^{0,0,0}_{l,l^\prime,L} C^{m,-m^\prime,M}_{l,l^\prime,L}\,.
\label{eq:coeffTranslateEff}
\end{equation}
Among the quantities recorded for the $i^{th}$ event in a data sample derived from Monte Carlo simulation  are the truth values of $\theta^*$ and $\phi^*$, and a discrete variable $s$ describing whether or not the generated event was accepted by the final analysis cuts. We let $s=1$ indicate that the event was accepted and $s=0$ indicate that it was rejected.   The probability for this to happen is 
\begin{equation}
        P_i(s;\theta^*,\phi^*,f_{l,m}) =  \delta_{1s}\epsilon(\theta^*, \phi^*; f_{l,m})  + 
 \delta_{0s}\left(1-\epsilon(\theta^*, \phi^*; f_{l,m})\right)\,.
\end{equation}
We can estimate the parameters of the efficiency by minimizing
\begin{equation}
-2\ln{\cal L} (f_{l,m})= -2 \sum_i \ln {\left (  \delta_{1s_i}\epsilon(\theta^*_i, \phi^*_i; f_{l,m}) + 
 \delta_{0s_i}(1-\epsilon(\theta^*_i, \phi^*_i;f_{l,m}) \right)}\,.
\label{effLogLik0}
\end{equation}
In case weighted events are used we incorporate the weight $w_i$ of the event as
\begin{equation}
-2\ln{\cal L}  (f_{l,m})= -2 \sum_i w_i \ln {\left (  \delta_{1s_i}\epsilon(\theta^*_i, \phi^*_i; f_{l,m}) + 
 \delta_{0s_i}(1-\epsilon(\theta^*_i, \phi^*_i;f_{l,m}) \right)}\,.
\label{effLogLik1}
\end{equation}
The coefficients $f_{l,m}$ and the overall efficiency factor $A$ can parametrized and fit as described in the previous 
section. 

\subsection {Angular resolution}
In this section we show how to obtain coefficients $r_l$ of an expansion
of the resolution function ${\cal R}(\Theta)$ in terms of Legendre polynomials:
\begin{equation}
{\cal R}(\Theta) =  r_l P_l(\cos{\Theta}).  
\end{equation}
This model assumes that the probability of a given deviation in measured lepton direction is universal, having no dependence 
on lepton direction and no anisotropy.\footnote{A more general model is discussed in Section~\ref{sec:xTended}.}\ Input to 
the fit is simulated data subject to the same reconstruction procedures as real data. For each event the true and reconstructed 
lepton directions in $\theta^*$, $\phi^*$ space are recorded and the angular deviation is computed.  
The coefficients are obtained by fitting a normalized, positive-definite fitting function to the distribution of angular deviations.
The fitting function is
\begin{equation}
{\cal R}(\Theta) = \left|u_l \sqrt{\frac{2l+1}{2}} \cdot P_l(\cos{\Theta})\right|^2
\end{equation}
where $u_l$ are complex coefficients satisfying 
\begin{equation}
\sum_l |u_l|^2 = 1\,.
\end{equation}
We parametrize these coefficients along the same lines as in previous
sections and then re-express the resolution function ${\cal R}(\Theta)$ as an
expansion in Legendre polynomials (rather than the square of such an
expansion), using a variant of Gaunt's theorem:
\begin{eqnarray}
{\cal R}(\Theta) & = & \left|u_l \sqrt{\frac{2l+1.0}{2.0}}P_l(\cos{\Theta})\right|^2 \nonumber \\
                & = & \frac {1}{2} u_l u^*_{l^\prime} \sqrt{(2l+1.0)(2l^\prime+1)} P_l(\cos{\Theta})P_{l^\prime}(\cos{\Theta}) \nonumber \\
                & = &  \sqrt{(2L+1)\pi} u_l u^*_{l^\prime} G_{l,l^\prime,L}^{0,0,0} P_L(\cos{\Theta})
\end{eqnarray}
which has the form:
\begin{equation}
{\cal R}(\Theta) =  r_L P_L(\cos{\Theta}),
\end{equation}
the $r_L$ being given by:
\begin{equation}
r_L = u_l u^*_{l^\prime} \sqrt{\frac{(2l+1)(2l^\prime + 1)}{4}} \left[C^{0,0,0}_{l,l^\prime,L}\right]^2
\label{eq:GauntLegendre}
\end{equation}
using again a Gaunt expansion for the product of two Legendre polynomials, expressed in terms of 
Clebsh-Gordan coefficients. \emph{Again}, we use the 
now-familiar strategy to fit a positive-definite, normalized function 
to our simulated data, and then (this time with eq.~\ref{eq:GauntLegendre}) transform to a more convenient 
form that can be used in eq.~\ref{eq:coeff}.

\section{Fit strategies}
The fractions $F_0$ of longitudinal $W$ bosons, $F_+$ of right-handed $W$ bosons,
and $F_-$ of left-handed $W$ bosons are commonly encountered in the literature on top quark couplings~\cite{ref:Kane}.  The relationship between these
helicity fractions and our parametrization is:
\begin{eqnarray}
f_1   & = & F_+ + F_- = 1 - F_0 \nonumber \\
f_1^+ & = & F_+/(F_+ + F_-) \,. 
\end{eqnarray}
SM values for the helicity fractions computed to next-to-next-to-leading order are~\cite{ref:theoryHelicity}:
\begin{eqnarray}
     F_+ &=& (1.7\pm0.1)\times 10^{-3}     \nonumber \\
     F_0 &=& 0.687\pm0.005  \nonumber \\
     F_- &=& 0.311\pm0.005\  
\end{eqnarray}
implying:
\begin{eqnarray}
f_1   & =  F_+ + F_- = 1 - F_0 = 0.313\pm0.005 \nonumber \\
f_1^+ & =  F_+/(F_+ + F_-) = .00543\pm0.00033 
\end{eqnarray}
while also in the SM we expect 
\begin{equation}
f_0^+ = {\cal O}\left(\frac{m_b^2}{m_W^2}\right)\,.
\end{equation}
These values have important consequences for strategies to extract these parameters from a dataset using likelihood techniques.  First, they imply that
at the SM point, $A_{1,1/2}$ and $A_{0,1/2}$ are both nearly zero.  The
fractions are practically on the edge of the physical parameter space, and the
likelihood becomes invariant to the parameter $\delta_+$.  Under these circumstances the double differential
decay rate, eq.~\ref{eq:doubleDiff1}, loses its sensitivity to the polarization
$P$ and the parameter $f_0^+$ separately, and instead depends only on the product
$P\sqrt{1-f_0^+}$. Since no other measurement presently determines $f_0^+$, we 
appear to be insensitive to a valuable piece of information near to the SM expected 
point.  

On the other hand \emph{we are still very sensitive to the most important physics parameter},
$\delta_-$, because the $P\sqrt{1-f_0^+}$ will now appear as the amplitude in the 
$\phi^*$-modulation, where $\delta_-$ plays the role of a phase.  To leading order in 
the anomalous couplings,
\begin{equation}
\delta_-=-\tan^{-1}\left[(x_W^{-1}-x_W){\rm Im}(g_R)\right]
\end{equation}
where $x_W=m_W/m_t$. Thus we can get a clean measurement of the most important physics quantity in the analysis, with no external dependence on the single top quark polarization.
 
Helicity fractions in top quark decay have been measured in $t{\bar t}$ production
by CDF(\cite{ref:cdfWHel1}-\cite{ref:cdfWHel8}) and by D\O\ (\cite{ref:d0WHel1}-\cite{ref:d0WHel5}) and combined~\cite{ref:comboWHel}; and more recently by ATLAS~\cite{ref:atlasWHel} and CMS~(\cite{ref:cmsWHel},\cite{ref:cmsWHelSingle}). Parameters extracted from single top quark 
$t$-channel events using techniques based upon the likelihood function described
here can be compared results from the $t{\bar t}$ analysis; consistency between the two measured values could provide strong indications that
the background is well understood and the signal is consistent with single top quark
production. Once the consistency was established, the measurements could be 
imported into the likelihood as an external constraint.

\section  {Extended models of detector effects}

The treatment of detector effects developed in the previous sections is based
on a model of detector effects that may, when used with real data, prove too
simple to describe all effects.  Namely, the model assumes that detector smearing is isotropic and does not depend on angle, and it also assumes that the 
efficiency function does not depend upon the physics parameters.  In this 
section we show how these effects can be properly handled within the same
framework of orthogonal polynomials, at the price of a small increase in the
complexity of the formulation. 

\subsection{Extended models of angular resolution}
\label{sec:xTended}

Let the true direction of a lepton in $\theta^*$, $\phi^*$ space be denoted by 
\begin{equation}
{\hat u} \equiv (\sin\theta_T^*\cos\phi^*_T,\sin\theta_T^*\sin\phi_T^*,\cos\theta_T^*)
\end{equation} 
and the reconstructed direction by
\begin{equation} 
{\hat v} \equiv (\sin\theta^*\cos\phi^*,\sin\theta^*\sin\phi^*,\cos\theta^*)\,.
\end{equation} 
As usual we express the underlying probability density as $f(\hat{u})=a_{l,m}Y_l^m(\theta^*_T, \phi^*_T)$. The probability that a lepton produced along 
${\hat u}$ will be reconstructed at $\hat{v}$ will be denoted as $g(\hat{u},\hat{v})$. The observed distribution $h(\hat{v})$ is then the integral
\begin{equation}
h(\hat{v}) = \int f(\hat{u})g(\hat{u},\hat{v}) d\Omega^*_T\,.
\end{equation}
The assumption of isotropic, angle-independent smearing is equivalent to the condition that 
\begin{equation}
g(\hat{u},\hat{v})=g({\hat u} \cdot {\hat v})\,,
\end{equation}
i.e. that the smearing probability only depends upon the angle between true
and reconstructed directions.  Under these circumstances, one can expand
\begin{equation}
g({\hat u} \cdot {\hat v}) = \frac{1}{2\pi} r_k P_k(\hat{u}\cdot\hat{v})
\end{equation}
and write
\begin{equation}
h(\hat{v}) = \frac{1}{2\pi} a_{l,m}r_k \int Y_l^m(\theta^*_T,\phi^*_T)  P_k(\hat{u}\cdot\hat{v}) d\Omega^*_T\,.
\end{equation}
Using the addition theorem, 
\begin{equation}
P_k(\hat{u}\cdot\hat{v}) = \frac{4\pi}{2k+1} \sum_{n=-k}^{k}Y_{k}^{n*}(\theta^*_T,\phi^*_T)Y_{k}^n(\theta^*,\phi^*)
\end{equation}
(no summation over $k$ on the right hand side), we have
\begin{eqnarray}
h(\hat{v}) &=& \frac{2}{2k+1} a_{l,m}r_k \int Y_l^m(\theta^*_T,\phi^*_T) Y_{k}^{n*}(\theta^*_T,\phi^*_T)Y_{k}^n(\theta^*,\phi^*)  d\Omega^*_T \nonumber \\
           &=&  \frac{2}{2k+1}a_{l,m}r_k \delta_{l,k}\delta_{m,n} Y_{k}^n(\theta^*,\phi^*) \nonumber \\
 &=&  d_{l,m} Y_{l}^m(\theta^*,\phi^*)
\end{eqnarray}
where 
\begin{equation}
d_{l,m}=\frac{2}{2l+1}a_{l,m}r_l
\end{equation}
(no summation over l) and one recovers the Funk-Hecke theorem, eqs.~\ref{eq:fh1} and~\ref{eq:fh2}. Now, if
\begin{equation}
g(\hat{u},\hat{v})\ne g({\hat u} \cdot {\hat v})
\end{equation}
then we can still expand the resolution function as
\begin{equation}
g(\hat{u},\hat{v})=r_{l,m,l^\prime,m^\prime}Y_{l}^m(\theta^*_T,\phi^*_T)Y_{l^\prime}^{m^\prime}(\theta^*,\phi^*)\,,
\end{equation}
in which case a few algebraic manipulations give us
\begin{equation}
h(\hat{v}) =  a_{l,-m}r_{l,m,l^\prime,m^\prime}(-1)^{m} Y_{l^\prime}^{m^\prime}(\theta^*,\phi^*)
\end{equation}
which is another type of convolution theorem for spherical harmonics, but with a completely general convolution kernel. 
Using this instead of the Funk-Hecke theorem modifies the coefficient formula, eq.~\ref{eq:coeffFormula}, so that it instead reads:
\begin{equation}
{\cal A}_{l,m}(\vec{\alpha})=\frac{G_{\lambda,\lambda^\prime,l^\prime}^{\mu, \mu^\prime, -m^\prime} a_{\lambda\mu}  (\vec{\alpha}) e_{\lambda^\prime \mu^\prime } r_{l^\prime,m^\prime,l,m}(-1)^{m^\prime}}{(-1)^{\sigma}  a_{\kappa,\sigma}(\vec{\alpha}) {e_{\kappa,-\sigma} }} 
+ (1-f_s)b_{l,m}\,.
\end{equation}
To find the coefficients $r_{l,m,l^\prime,m^\prime}$ we construct a positive-definite normalized fitting function:
\begin{equation}
g(\hat{u},\hat{v})={\cal R}(\theta^*_T, \phi^*_T,\theta^*,\phi^*) = \left |\sum_{\lambda,\mu,\kappa,\nu} s_{\lambda,\mu,\kappa,\nu}Y_\lambda^\mu(\theta^*_T, \phi^*_T)Y_\kappa^\nu(\theta^*, \phi^*)\right|^2      
\label{eq:bkgDeffXt}
\end{equation}
where the $s_{\lambda,\mu,\kappa,\nu}$ are complex coefficients satisfying 
\begin{equation}
\sum_{\lambda,\mu,\kappa,\nu} |s_{\lambda,\mu,\kappa,\nu}|^2 = 1\,.
\label{eq:normDefXt}
\end{equation}
The constants $s_{\lambda,\mu,\kappa,\nu}$ can be determined in a likelihood fit to signal Monte Carlo. Then, the $r_{l,m,l^\prime,m^\prime}$ are obtained from the $s_{\lambda,\mu,\kappa,\nu}$ via the relationship
\begin{eqnarray}
r_{l,m,l^\prime,m^\prime} &=&  (-1)^{\mu^{\prime}+\nu^{\prime}}G^{\mu,-\mu^\prime,m}_{\lambda, \lambda^\prime, l}G^{\nu,-\nu^\prime,m^\prime}_{\kappa, \kappa^\prime, l^\prime} s_{\lambda,\mu,\kappa,\nu}s^*_{\lambda^\prime,\mu^\prime,\kappa^\prime,\nu^\prime}\,. 
\label{eq:coeffTranslateXt}
\end{eqnarray}

\subsection{Variation of the Efficiency}

It can happen that the efficiency, $\epsilon(\theta^*, \phi^*) = e_{l,m}Y_l^m(\theta^*,\phi^*)$, varies as a function of the physics parameters $\vec{\alpha}$:
\begin{equation}
\epsilon(\theta^*, \phi^*; \vec{\alpha}) = e_{l,m}(\vec{\alpha})Y_l^m(\theta^*,\phi^*)\,.
\label{eq:effvariation}
\end{equation}
This arises because of the integration over the angle $\theta$, which we carried out to obtain eq.~\ref{eq:doubleDiff1}. 
The efficiency as a function of \emph{three} observable angles $\varepsilon(\theta,\theta^*,\phi^*)$, when projected into 
only two angles ($\theta^*$ and $\phi^*$), acquires a dependence on $\vec{\alpha}$, as follows:
\begin{eqnarray}
\epsilon(\theta^*,\phi^*; \vec{\alpha})=\frac{\int \varrho(\theta,\theta^*,\phi^*;\vec{\alpha})\varepsilon(\theta,\theta^*,\phi^*)d(\cos\theta)}{\int\varrho(\theta,\theta^*,\phi^*)d(\cos\theta)}\,.
\end{eqnarray}
This dependency can be accounted for with a slight modification of the formalism
developed here, without sacrificing any of the mathematical elegance.  We can now express the triple differential decay rate, eq.~\ref{eq:trippleDiff} as
\begin{equation}          
   \varrho(\theta,\theta^*,\phi^*) = s_{k,n,l,m}(\vec{\alpha})P_k^n(\cos\theta) Y_l^m(\theta^*,\phi^*)
\end{equation}
where:
\begin{equation}
\begin{split}
&s_{0,0,0,0}  = \frac{ a_{0,0}}{4\pi} \,,   \qquad     s_{0,0,1,0}  = \frac{a_{1,0}}{4\pi} \,, \qquad s_{0,0,2,0} = \frac{a_{2,0}}{4\pi}    \,,   \qquad     s_{1,1,1,-1}  = -\pi^2  a_{1,-1} \,, \\
&s_{1,1,1,1}  = -\pi^2  a_{1,1}\,, \qquad s_{1,1,2,-1}   = -\pi^2  a_{2,-1} \,, \qquad s_{1,1,2,1}  =-\pi^2 a_{2,1}\,,  \\
&s_{1,0,0,0}  = \frac{P}{(4\pi)^{3/2}} \left(f_1\left(2f_1^+-1\right) - \left(1-f_1\right)\left(2f_0^+-1 \right)\right) \,, \qquad s_{1,0,1,0}  =  \frac{P\sqrt{3}}{2(4\pi)^{3/2}}f_1\,, \\
&s_{1,0,2,0}  =  \frac{P}{\sqrt{5}(4\pi)^{3/2}}\left(\frac{1}{2}f_1\left(2f_1^+-1\right) + \left(1-f_1\right)\left(2f_0^+-1 \right)\right) \\
\end{split}
\label{eq:physicsCoefficients3D}
\end{equation}
and all of the other coefficients are zero. Expanding the efficiency function,
\begin{equation}
\varepsilon(\theta,\theta^*,\phi^*)=w_{k,l,m}P_k(\cos\theta)Y_l^m(\theta^*,\phi^*)\,,
\end{equation}
and integrating over $\cos\theta$ gives us:
\begin{eqnarray}
\epsilon(\theta^*,\phi^*; \vec{\alpha}) &=&\frac{\int \varrho(\theta,\theta^*,\phi^*;\vec{\alpha})\varepsilon(\theta,\theta^*,\phi^*)d(\cos\theta)}{\int\varrho(\theta,\theta^*,\phi^*)d(\cos\theta)} \nonumber \\
                         & = & \frac
{\int s_{\kappa,\nu,\lambda,\mu}(\vec{\alpha})P_{\kappa}^{\nu}(\cos\theta)Y_{\lambda}^{\mu}(\theta^*,\phi^*)\cdot w_{k,l,m}P_k(\cos\theta)Y_l^m(\theta^*,\phi^*)d(\cos\theta)}
{a_{\lambda,\mu}(\vec{\alpha})Y_{\lambda}^{\mu}(\theta^*,\phi^*)} \nonumber \\
                         & = & \frac
{\int s_{\kappa,\nu,\lambda,\mu}(\vec{\alpha})P_{\kappa}^{\nu}(\cos\theta)Y_{\lambda}^{\mu}(\theta^*,\phi^*)\cdot w_{k,l,m}P_k(\cos\theta)Y_l^m(\theta^*,\phi^*)d(\cos\theta)}
{a_{\lambda,\mu}(\vec{\alpha})Y_{\lambda}^{\mu}(\theta^*,\phi^*)}
\nonumber \\
                         & = & \frac
{s_{\kappa,\nu,\lambda,\mu}(\vec{\alpha})w_{k,l,m} Y_{\lambda}^{\mu}(\theta^*,\phi^*)  Y_l^m(\theta^*,\phi^*)}
{a_{\lambda,\mu}(\vec{\alpha})Y_{\lambda}^{\mu}(\theta^*,\phi^*)}\int P_k(\cos\theta) P_{\kappa}^{\nu}(\cos\theta) d(\cos\theta)\nonumber \\ 
                         & = & \frac
{ S_{k,\kappa}^{\nu} s_{\kappa,\nu,\lambda,\mu}(\vec{\alpha})w_{k,l,m} Y_{\lambda}^{\mu}(\theta^*,\phi^*)  Y_l^m(\theta^*,\phi^*)}
{a_{\lambda,\mu}(\vec{\alpha})Y_{\lambda}^{\mu}(\theta^*,\phi^*)}\nonumber \\
                         & = & \frac
{ G_{\lambda, l, L}^{\mu,m,M} S_{k,\kappa}^{\nu} s_{\kappa,\nu,\lambda,\mu}(\vec{\alpha})w_{k,l,m} Y_{L}^{M}(\theta^*,\phi^*)}
{a_{\lambda,\mu}(\vec{\alpha})Y_{\lambda}^{\mu}(\theta^*,\phi^*)}\,, 
\label{eq:effVaryDeriv}
\end{eqnarray}
where we have introduced the constant 
\begin{equation}
 S_{k,\kappa}^{\nu}\equiv \int_{-1}^{1} P_k(x) P_{\kappa}^{\nu}(x) dx
\end{equation}
An analytic expression for this integral can be easily derived with the use of Gaunt's theorem:
\begin{equation}
 S_{k,\kappa}^{\nu} = \sqrt{\frac{4\pi(2L+1)}{(2k+1)(2\kappa+1)}\frac{(\kappa+\nu)!}{(\kappa-\nu)!}\frac{(L-\nu)!}{(L+\nu)!}}G^{0,\nu,\nu}_{k,\kappa,L} R_L^M
\end{equation} 
where 
\begin{equation}
R_L^M \equiv \int_{-1}^{1} P_L^M(x) dx
\end{equation}
is the integral of the associated Legendre function. Explicit expressions for this integral can be found in ref.~\cite{ref:AssInt}. 
Our goal here is to obtain an expression for the coefficients $e_{L,M}(\vec{\alpha})$; we can achieve this by equating expansion~\ref{eq:effVaryDeriv} to the expansion ~\ref{eq:effvariation}:
\begin{eqnarray}
\epsilon(\theta^*,\phi^*; \vec{\alpha}) &=&  \frac
{ G_{\lambda, l, L}^{\mu,m,M} S_{k,\kappa}^{\nu} s_{\kappa,\nu,\lambda,\mu}(\vec{\alpha})w_{k,l,m} Y_{L}^{M}(\theta^*,\phi^*)}
{a_{\lambda,\mu}(\vec{\alpha})Y_{\lambda}^{\mu}(\theta^*,\phi^*)} \nonumber \\
                          &=& e_{l,m} (\vec{\alpha}) Y_{l}^{m}(\theta^*,\phi^*) 
\end{eqnarray}
so 
\begin{eqnarray}
{ G_{\lambda, l, L}^{\mu,m,M} S_{k,\kappa}^{\nu} s_{\kappa,\nu,\lambda,\mu}(\vec{\alpha})w_{k,l,m} Y_{L}^{M}(\theta^*,\phi^*)}
&=& {a_{\lambda,\mu}(\vec{\alpha}) e_{l,m}(\vec{\alpha}) Y_{\lambda}^{\mu}(\theta^*,\phi^*)}Y_{l}^{m}(\theta^*,\phi^*) \nonumber \\
&=& {a_{\lambda,\mu}(\vec{\alpha}) e_{l,m}(\vec{\alpha}) G_{\lambda,l,L}^{\mu,m,M}Y_{L}^{M}(\theta^*,\phi^*)} 
\end{eqnarray}
from which we can conclude that:
\begin{eqnarray}
S_{k,\kappa}^{\nu}w_{k,l,m}s_{\kappa,\nu,\lambda,\mu} (\vec{\alpha}) &=& {a_{\lambda,\mu}(\vec{\alpha}) e_{l,m}(\vec{\alpha}) }\,, 
\end{eqnarray}
or
\begin{eqnarray}
e_{l,m}(\vec{\alpha}) &=& \frac {S_{k,\kappa}^{\nu}w_{k,l,m}s_{\kappa,\nu,\lambda,\mu} (\vec{\alpha})}  {a_{\lambda,\mu}(\vec{\alpha})}\,.
\label{eq:effcoeffvary}
\end{eqnarray}
This expression now contains the variation of the efficiency parameters as a function of physics parameters $\vec{\alpha}$.
Eq.~\ref{eq:coeff} can now be used without modification, provided that the coefficients in eq.~\ref{eq:effcoeffvary} are 
taken instead of fixed coefficients.

This leaves us with the problem of determining the coefficients $w_{k,l,m}$.  This can be carried out in a straightforward extension of the methods previously discussed.  First, we construct a parametrized efficiency function which is positive-definite:
\begin{equation}
\epsilon \left ( \theta ^{\ast },\varphi^{\ast},\theta\right )=A^{2}\cdot \left | \sum_{k,l,m} t_{k,l,m}P_{k}(\cos\theta )Y_{l}^{m}(\theta ^{\ast} ,\varphi ^{\ast }) \right |^{2}
\end{equation}
where the $t_{k,l,m}$ are complex coefficients satisfying 
\begin{equation}
\sum_{k,l,m} |t_{k,l,m}|^2 = 1\,.
\end{equation}
The coefficients $t_{k,l,m}$ can be determined from a likelihood fit to signal Monte Carlo. Then we write 
\begin{eqnarray}
\epsilon ( \theta ^*,\varphi^*,\theta )=A^{2}\cdot t_{k,l,m}P_{k}Y_{l}^{m}t_{k^\prime, l^\prime, m^\prime}^*P^*_{k^\prime}Y_{l^\prime}^{-m^\prime}(-1)^{m^\prime} \,.
\end{eqnarray}
Applying Gaunt's theorem
\begin{equation}
P_{n}(x)P_{n^\prime}(x)=\sum_{m=0}^{n+n^\prime}G_{m}^{n,n^\prime}P_{m}(x)\,,
\end{equation}
\begin{equation}
Y_{l}^{m}( \theta ,\varphi)Y_{l^\prime}^{m^\prime}( \theta ,\varphi)=\sum_{L,M}G_{l,l^\prime, L}^{m,m^\prime, M}Y_{L}^{M}\left ( \theta ,\varphi  \right)\,,
\end{equation}
where
\begin{equation}
  G_{m}^{n,n^\prime} =\left[C_{n,n^{\prime},m}^{0,0,0}\right]^2
\end{equation}
gives us
\begin{eqnarray}
\epsilon(  \theta^{*}, \varphi ^{*},\theta) &=& A^{2}\cdot t_{k,l,m}t_{k^\prime, l^\prime, m^\prime}^*G_{i}^{k,k^\prime}P_{i}(\cos\theta )G_{l,l^\prime, L}^{m,{-m^\prime},M}Y_{L}^{M}\left ( -1 \right )^{m^\prime}  \nonumber \\
&=&A^{2}\cdot (-1)^{m^\prime}t_{k,l,m}t{_{k^\prime, l^\prime, m^\prime}^*}G_{i}^{k,k^\prime}G_{l,l^\prime, L}^{m,{-m^\prime}, M}P_{i}(\cos\theta )Y_{L}^{M}(\theta^*, \varphi^* )\,, 
\end{eqnarray}
so that the translation from the $t_{k,l,m}$ to the $w_{k,l,m}$ is via the relation:
\begin{equation}
w_{i,L,M}=A^2 \cdot (-1)^{m^\prime} t_{k,l,m}t^*_{k^\prime, l^\prime, m^\prime}G_i^{k,k^\prime}G_{l,l^\prime,L}^{m,{-m^\prime},M}\,.
\end{equation}

\section{Conclusions}

We have presented here formulae for the analysis of the decay of polarized
single top quarks produced in the $t$-channel.  The formulae can be
used in a variety of likelihood-based techniques to precisely measure
the $Wtb$ vertex from angular distributions measured at the LHC. 
While formulae governing the raw physics PDFs are easy to obtain, the 
main difficulty arises in the convolution of detector effects.  In the 
past only Monte Carlo simulation has been capable of addressing this 
aspect, and full simulation is too time consuming to be used at each step
in a   likelihood fit by many orders of magnitude. 

A theorem called the Funk-Hecke theorem, related to the convolution
theorem but applicable to an angular analysis in spherical
harmonics rather than a Fourier transform, is the basis of a practical
scheme for convolving detector effects with physics distributions. This
scheme, which we outline in this paper, requires Monte Carlo for a 
description of resolution and efficiency, then uses the 
formula~\ref{eq:coeff} or simple extensions to carry out the 
convolution.  

We have applied this to the double-differential decay rate in $t$-channel
single top quark decay, described by eq.~\ref{eq:doubleDiff1}.  This is
made possible by the existence of an angular decomposition of the differential
decay rate on one hand, and the properties of orthogonal polynomials on
the other.  Because these properties can be discovered in many interesting
systems, it is likely, that the basic technique can be applied 
elsewhere.  The most interesting systems are those where significant 
measurement errors are present, coming from the presence of such objects
as jets or a neutrino in the final state.



\appendix
\section{Analytic minimization}
In this appendix we show that we can analytically minimize
the likelihood function for several of our fits when the PDF is expressed
in terms of spherical harmonics --- this is the way, for example, that
track fits are normally carried out in a reconstruction program. Since
the computations are long we will illustrate this only for the simplest 
type of fit, namely the kind we use for the background shape, where the 
likelihood function is:

\begin{equation}
-2ln\mathcal{L}=\sum_{j}w_{j}\ln(|g_{l,m}y_{l,m}^{j}|^{2})
\end{equation}
where the $g_{l,m}$ are complex coefficients.  We decompose these into
real and imaginary parts
\begin{equation}
g_{l,m}=a_{l,m}+ib_{l,m}
\end{equation}
and there are two real parameters for every complex coefficient. Minimizing
with respect to the real part $a_{l,m}$ of the coefficients gives
us 
\begin{eqnarray}
\frac{\partial(-2\ln{\cal L})}{\partial a_{L,M}} & = & -2\sum_{j}\frac{w_{j}}{|g_{r,s}y_{r,s}^{j}|^{2}}\frac{\partial}{\partial a_{L,M}}\left(|g_{l,m}y_{l,m}^{j}|^{2}\right)\,.
\end{eqnarray}
After some straightforward calculation we obtain:
\begin{eqnarray}
\frac{\partial(-2\ln{\cal L})}{\partial a_{L,M}} & = & -4{\rm Re}\left[\sum_{j}w_{j}\frac{y_{L,M}^{j}}{g_{l,m}y_{l,m}^{j}}\right]\,, \nonumber \\
\frac{\partial(-2\ln{\cal L})}{\partial b_{L,M}} & = &  4{\rm Im}\left[\sum_{j}w_{j}\frac{y_{L,M}^{j}}{g_{l,m}y_{l,m}^{j}}\right]\,. 
\end{eqnarray}
Next, we require second derivatives, in particular we need the full
$(L_{max}+1)^{2}\times(L_{max}+1)^{2}$ matrix:
\[
\left(\begin{array}{cc}
\frac{\partial^{2}(-2\ln{\cal L})}{\partial a_{\lambda,\mu}\partial a_{L,M}} & \frac{\partial^{2}(-2\ln{\cal L})}{\partial a_{\lambda,\mu}\partial b_{L,M}}\\
\frac{\partial^{2}(-2\ln{\cal L})}{\partial b_{\lambda,\mu}\partial a_{L,M}} & \frac{\partial^{2}(-2\ln{\cal L})}{\partial b_{\lambda,\mu}\partial b_{L,M}}
\end{array}\right),
\]
which is 
\begin{eqnarray}
\frac{\partial^{2}(-2\ln{\cal L})}{\partial a_{R,S}\partial a_{L,M}} & = & \frac{\partial}{\partial a_{R,S}}\left(-2\sum_{j}w_{j}\left[\frac{y_{L,M}^{j}}{g_{l,m}y_{l,m}^{j}}+CC\right]\right)\nonumber \\
 & = & +4{\rm Re}\sum_{j}w_{j}\left[\frac{y_{L,M}^{j}y_{R,S}^{j}}{(g_{l,m}y_{l,m}^{j})^{2}}\right]\,.
\end{eqnarray}
Likewise,
\begin{eqnarray}
\frac{\partial^{2}(-2\ln{\cal L})}{\partial b_{R,S}\partial a_{L,M}} & = & \frac{\partial}{\partial b_{R,S}}\left(-2\sum_{j}w_{j}\left[\frac{y_{L,M}^{j}}{g_{l,m}y_{l,m}^{j}}+CC\right]\right)\nonumber \\
 & = & -4{\rm Im}\sum_{j}w_{j}\left[\frac{y_{L,M}^{j}y_{R,S}^{j}}{(g_{l,m}y_{l,m}^{j})^{2}}\right]\,.
\end{eqnarray}
and
\begin{eqnarray}
\frac{\partial^{2}(-2\ln{\cal L})}{\partial b_{R,S}\partial b_{L,M}} & = & \frac{\partial}{\partial b_{R,S}}\left(-2\sum_{j}w_{j}\left[i\frac{y_{L,M}^{j}}{g_{l,m}y_{l,m}^{j}}+CC\right]\right)\nonumber \\
 & = & -4{\rm Re}\sum_{j}w_{j}\left[\frac{y_{L,M}^{j}y_{R,S}^{j}}{(g_{l,m}y_{l,m}^{j})^{2}}\right]\,.
\end{eqnarray}
To minimize the likelihood, one first assigns random values to the coupling 
$g_{l,m}.$ Then, one computes the sums 
\begin{equation}
S_1=\sum_{j}w_{j}\frac{y_{L,M}^{j}}{g_{l,m}y_{l,m}^{j}}\qquad S_2=\sum_{j}w_{j}\left[\frac{y_{L,M}^{j}y_{R,S}^{j}}{(g_{l,m}y_{l,m}^{j})^{2}}\right]\,.
\end{equation}
The first of these (real and imaginary parts) is used to compute the
first derivative of the likelihood, the second is used to compute
its second derivative. Taking the real and imaginary parts of the
full coefficient set $\{g_{l,m}\}$ as a large, real-valued vector
$\vec{\eta}$ and denoting
\begin{eqnarray}
\frac{\partial(-2\ln{\cal L})}{\partial\vec{\eta}} & \equiv & \vec{\xi}\,,\nonumber\\
\frac{\partial(-2\ln{\cal L})}{\partial\vec{\eta}\partial\vec{\eta}} & \equiv & \mathbf{\mathbf{H}}
\end{eqnarray}
 we have:
\begin{equation}
\mathbf{H}\left(\vec{\eta}_{0}-\vec{\eta}\right)=-\vec{\xi}\label{eq:MatrixMethod}
\end{equation}
which relates the parameters which minimize the likelihood, $\vec{\eta}_{0}$,
to starting parameters $\vec{\eta}$ and starting derivatives $\vec{\xi}$.
The solution is 
\begin{equation}
\vec{\eta}_{0}=\mathbf{H}^{-1}\cdot\left(\mathbf{H}\cdot\vec{\eta}-\vec{\xi}\right)\,.
\label{eq:anal}
\end{equation}
One can test this by computing the derivatives of the likelihood function,
$\vec{\xi},$ again at the solution to this equation, as well as the
Hessian $\mathbf{H}$; if the first step has not succeeded in locating
bringing the fit to a true minimum (because eq.~\ref{eq:MatrixMethod} ignores
higher order terms) the procedure can be iterated, and when convergence
has occurred both the parameter set $\vec{\eta}_{0}$ and the Hessian
can be saved, the latter being useful to compute the covariance matrix
from the relation $\mathbf{C}^{-1}=\mathbf{H}/2$. For the background
fit, some special considerations also apply. First, the likelihood
function is insensitive to an overall phase. This degree of freedom
can be removed by fixing the phase of $g_{0,0}$ to zero. We also have to 
deal with the normalization constraint, $g_{l,m}g_{l,m}^{*}=1$. We do this by
enlarging the set of parameters with a Lagrange multiplier $\lambda$,
and modifying the likelihood correspondingly:
\begin{equation}
-2\ln{\cal L}\rightarrow\Lambda=-2\ln{\cal L}+\lambda(g_{l,m}g_{l,m}^{*}-1)\,.
\end{equation}
The extra parameter adds another element to the vector $\vec{\xi}$
which is 
\begin{equation}
\frac{\partial\Lambda}{\partial\lambda}=(g_{l,m}g_{l,m}^{*}-1)\,.
\end{equation}
The second derivative is zero:
\begin{equation}
\frac{\partial^{2}\Lambda}{\partial\lambda^{2}}=0
\end{equation}
but the extra terms modify the derivatives according to 
\begin{eqnarray}
\frac{\partial\Lambda}{\partial a_{L,M}} & = & \frac{\partial\left(-2\ln{\cal L}\right)}{\partial a_{L,M}}+\lambda(g_{L,M}^{*}+g_{L,M})\,,\nonumber\\
\frac{\partial\Lambda}{\partial b_{L,M}} & = & \frac{\partial\left(-2\ln{\cal L}\right)}{\partial b_{L,M}}+\lambda(ig_{L,M}^{*}-ig_{L,M}),
\end{eqnarray}
and the Hessian $\mathbf{H}$ according to
\begin{eqnarray}
\frac{\partial^{2}\Lambda}{\partial a_{R,S}\partial a_{L,M}} & = & \frac{\partial^{2}\left(-2\ln{\cal L}\right)}{\partial a_{R,S}\partial a_{L,M}}+2\lambda\delta_{R,L}\delta_{S,M}\nonumber\\
\frac{\partial^{2}\Lambda}{\partial b_{R,S}\partial b_{L,M}} & = & \frac{\partial^{2}\left(-2\ln{\cal L}\right)}{\partial b_{R,S}\partial b_{L,M}}+2\lambda\delta_{R,L}\delta_{S,M}\nonumber\\
\frac{\partial^{2}\Lambda}{\partial b_{R,S}\partial a_{L,M}} & = & \frac{\partial^{2}\left(-2\ln{\cal L}\right)}{\partial b_{R,S}\partial a_{L,M}}\nonumber \\
\frac{\partial^{2}\Lambda}{\partial\lambda\partial a_{L,M}} & = & 2{\rm Re} g_{L,M}\nonumber\\
\frac{\partial^{2}\Lambda}{\partial\lambda\partial b_{L,M}} & = & 2{\rm Im} g_{L,M}
\end{eqnarray}
One can now proceed with analytic minimization as in eq.~\ref{eq:anal} using 
the enlarged set of parameters and the enlarged Hessian matrix.

\end{document}